\documentclass[aps]{article}
\usepackage{geometry}                
\usepackage{graphicx}

\usepackage{amssymb}
\usepackage{epstopdf}
\usepackage{amsmath}

\usepackage{amsfonts}
\usepackage{bm}

\DeclareMathAlphabet{\mathpzc}{OT1}{pzc}{m}{it}

\title{The Absence of Stokes Drift in Waves}
\author{Clifford Chafin\\\ \small{Department of Physics, North Carolina State University, Raleigh, NC 27695} \thanks{cechafin@ncsu.edu}}

\begin{document}
\maketitle
\begin{abstract}
Stokes drift has been as central to the history of wave theory as it has been distressingly absent from experiment.  Neither wave tanks nor experiments in open bodies detect this without nearly canceling ``eulerian flows.''  Acoustic waves have an analogous problem that is particularly problematic in the vorticity production at the edges of beams.  Here we demonstrate that the explanation for this arises from subtle end-of-packet and wavetrain gradient effects such as microbreaking events and wave-flow decomposition subtleties required to conserve mass and momentum and avoid fictitious external forces.  These losses occur at both ends of packets and can produce a significant nonviscous energy loss for translating and spreading surface wave packets and wavetrains.  In contrast, monochromatic sound wave packets will be shown to asymmetrically distort to conserve momentum.  This provides an interesting analogy to how such internal forces arise for gradients of electromagnetic wavetrains in media.  Such examples show that the interactions of waves in media are so system dependent  as to be completely nonuniversal.  These give further examples of how boundary effects must be carefully considered for conservation laws especially when harmonic functions are involved.  The induced flows in establishing surface waves are shown to be time changing and dependent on wave history and suggest that some classical work based on mass flux and wave interactions may need to be reconsidered.  
\end{abstract}

\section{Introduction}
The theory of waves in media has been an important spawning and testing ground for many of the topics of modern mathematical physics.  Nonlinear dynamics, perturbation theory and many methods of partial differential equations and vector calculus have electrodynamics and hydrodynamics as historical starting points and important introductory examples.  Hydrodynamics is particularly important because the nonlinearity in the system is so essential.  It is not the result of some dissipation but the fact that an eulerian description requires an advective term.  Since numerical methods were not available until the mid twentieth century, analytic methods and experiments were the only means of deriving intuition and understanding of these systems.  Exact nonlinear solutions are relatively rare but it was exactly such a case that was the first surface wave solution.  In the case of Gerstner waves, the particles exhibit circular orbits and remain on isobars \cite{Gerstner, Kalisch}.  It was later realized that these solutions have vorticity and are therefore not realistic wave solutions for propagating disturbances.  The first approximate solutions for small waves were due to Airy and Stokes \cite{Airy, Stokes} about 50 years later.  

It was noticed early on that these propagating periodic solutions, unlike the Gerstner case, led to a slow advance of the particles when their lagrangian motion was calculated.  Refinement of the solutions by perturbations showed that this was not an artifact of low order results and convergence of the Stokes series was shown by Levi-Civita so seemed to be an essential feature of ocean waves.  This meant the waves possessed a net momentum, a  concept that has been considered important in wave-flow interactions, wave-wave interactions and the set-up of waves on beaches.  Much of the elegant mathematical work since the 1950's elaborated on this.  

One of the difficult aspects of experimental confirmation of theory is that waves in nature are messy and often combined with surface flow, turbulence, wind and varying bottom geometry.  Even wave tank experiments \cite{Bagnold} often have to contend with the turbulence of the driver and damping surfaces and the fact that they were finite meant that there was a problem with continued mass flux at the ends.  The low viscosity of water, the primary fluid for experiment, meant that laminar results were elusive and so poor agreement with experiments could be dismissed.  Recent high precision data \cite{Smith, Monismith} has shown that inconsistencies in the drift predictions persist.  Wave tank results always need the drift removed to make any comparison with the higher order Stokes and cnoidal wave approximations.  Interpretation of this data has always been characterized by assuming the established wave theory is correct and we are missing an explanation for compensating ``eulerian flows'' or involve some vague and, often poorly considered, discussion of lagrangian mean flows.  

Recently, I have done a series of investigations on the implications of this drift on wavemaker theory and wave interactions to check for consistency with established results \cite{Chafin-wavemaker}.  Consistency with the literature was mixed as was expected.  Many of the classical results use ``momentum flux'' in a naive fashion and boundary conditions and conservation laws are inadequately checked.  Examples like this have been a persistent bane of electrodynamics regarding the hidden momentum \cite{Chafin-em}.  This was one of the motivations for this program.  Personal experiments and observations done over the years sought some validation of the existence of Stokes drift and thus the implied momentum of surface waves.  It is certainly true the rotating fluids in levitating traps with irrotational motion can have angular momentum.  This is seen in ultracold gas experiments \cite{Stringari}.  However, the generation of these rotating clouds typically involves further evaporation to allow particles with angular momentum to escape.  Simply rotating the harmonic trap does not automatically transfer angular momentum to the cloud and questions surrounding the role of the ``halo'' of excited particles remain to be answered.  Indeed, assuming pure irrotational flow to arbitrary distance creates a number of paradoxes when it comes to damping \cite{Chafin-thesis}.  

In large bodies of water with long fetch for wave growth there can be deep persistent currents driven by friction of the wind.  However in ponds and small lakes one can observe wave growth from one side to another.  Gusts can be irregular and drive small wave growth then die for appreciable periods of time.  The surface shear confined to the first centimeter of water will dissipate quickly into the bulk and leave little net motion.  I have observed in a wide variety of such environments from pollen and dust motion that the net drift imparted to the water from the waves seems to be exactly zero.  In many cases the net surface motion was backwards!  Never once, in such an environment, did the net surface motion advance for any more than a brief forwards surge after a gust.  These observations were usually in bodies large enough so there would be very little cost to large scale recirculation. 

There is an analogous example in the case of acoustics.  Acoustic streaming or Eckart streaming occurs when we have MHz frequency waves in a liquid that lead to an initially fast moving jet from the driver into the bulk of the fluid that then recirculates.  Lagrangian and nonlinear analysis of particles in acoustic waves leads to a net mass advance \cite{LL, Lighthill}.  Eventually, there is a steady recirculating flow that is maintained by the action of the sound.  Often it is implied that the net motion of the nonlinear sound wave has a role in driving this flow.  Given the nature of the transients and that, initially there is no flow at all, puts this into some doubt.  

In both the surface wave and acoustic wave case, vorticity conservation is a useful foil for such notions.  For all but the longest surface waves, the mass density is constant.  In acoustics, the incompressibility assumption must fail for sound to exist but it does so on very short time scales that average out so that vorticity is still generally well conserved on longer time scales than the oscillation time.  We will see that this gives a modification of the pressure field (and not true ``stress'') that is rapidly cancelled by elastic changes in the medium.  

This paper will begin with a quick review of Stokes drift for surface and acoustic waves.  Following this will be a discussion on the importance of packets and gradient based analyses with conserved quantities to get consistent results to contrast with infinite periodic solutions.  A specific example of EM waves in dielectrics illustrates this and shows how the ``ring up'' preparation of the medium for an advancing front can play an important role hidden from plane wave analyses.  Surface waves are examined first with surface constraints that are not physical, in that they rely on spurious forces, but help reproduce some of the previous (erroneous) notions of how mass transfer occurs in waves.  Relaxing these conditions we see that microbreaking is essential to conserve mass and momentum.  This eliminates all net mass flux across the surface of a resting body of water as a wave packet crosses.  The damping due to this is estimated and compared with viscous losses.  It is shown that a universal result exists for damping as a function of the oscillation of the envelope governing the wave.  

Acoustic wave momentum is considered and contrasted next.  Interestingly, the momentum conservation at wave fronts is not lossy as in the surface wave case but the result of a nonlinearity that is best understood by a basis of solutions that are consistent with boundary conditions.  Conditions on stationary packets are introduced along with the necessary shape changes to give a net zero momentum density as the waves propagate.  Such a condition is essential for hydrodynamic waves without fictitious forces on them.  These packets do not damp as they propagate in some nonviscous manner\footnote{I have refrained from using ``inviscid'' here since the damping of the waves is not related to the presence of internal viscous damping but wave breaking.}  as in the surface wave case but they do illustrate how medium changes outside of the support of the waves is essentially connected to such motion.  These elastic and kinetic changes provide a source of ring-up energy that acts as a hidden sink for any advancing wavetrain.  Such effects seem outside the range of consideration for the general lagrangian mean (GLM) approaches that seek to resolve the problem of net mass drift.

\section{Stokes Drift}
\subsection{Surface Waves}
\begin{figure}
  \begin{centering}
 \includegraphics[width=3in,trim=20mm 0 10mm 0mm,clip]{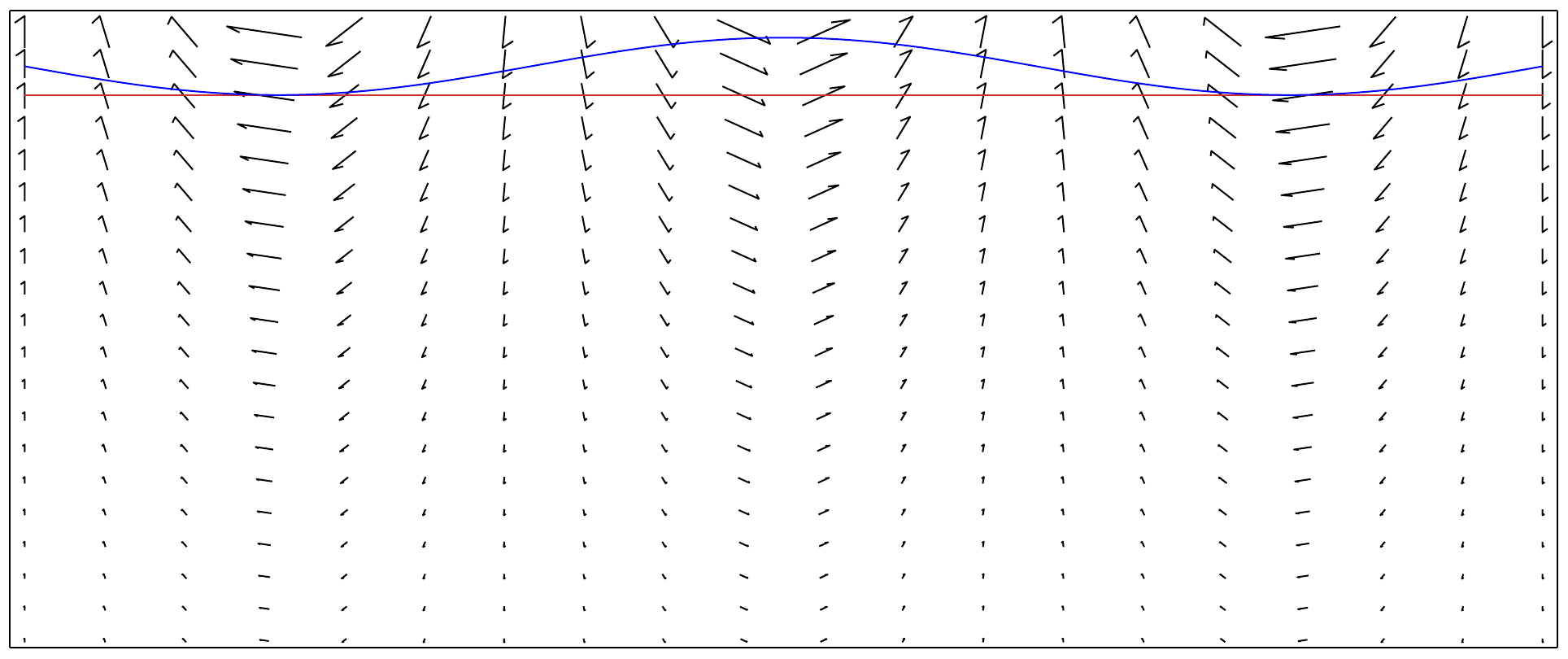}
 \caption{\label{wavecut} The irrotational velocity field superimposed on the elevation profile of an Airy wave.}
 \end{centering}
 \end{figure}

The linear limit of surface waves is more delicate than one would naively expect.  The case where the amplitude of the wave is much smaller than the wavelength gives a higher order contribution from the nonlinear term.  The periodic such solutions are called Airy waves.  The velocity field is superimposed on the wave profile as in fig.\ \ref{wavecut}.  The kinematic surface wave condition reads
\begin{equation}
\partial_{x}\Phi+\partial_{t}\Phi\partial_{x}\eta=\partial_{t}\eta
\end{equation}
so we see that there is a higher order mass flux source and sink on opposite faces of the wave.  The circulation of this flux through the wave is the Stokes drift.  Ideally we would impose this constraint exactly at each order in the perturbation series and seek corrections to surface shape and velocity fields that address both the nonlinear surface condition and the velocity potential equation
\begin{align}
\partial_{t}\Phi+\frac{1}{2}\nabla\Phi\cdot\nabla\Phi=-\frac{1}{\rho} P+ g  z
\end{align}
where constant density $\rho$ is assumed.  Pressure is determined in the case of waves that move much slower than the sound speed of the medium by 
\begin{align}
\nabla^{2}P=-\rho(\partial_{i}v_{j})^{2}=-(\partial_{ij}\phi)^{2}
\end{align}
with zero motion at depth and constant pressure surface boundary conditions.  The dynamic boundary condition is just a restatement that $P(x,\eta(x),t)$ is constant.  

The Airy solutions are given by
\begin{align}
\label{Airy}
\eta&=a \cos(k x- \omega t)\\
\Phi&=\frac{a \omega}{k} \sin(k x-\omega t)e^{kz}
\end{align}
for infinite depth waves of height $a$ and with dispersion law $\omega^{2}=gk$.  The perturbative assumption is that we can iteratively correct our way to the nonlinear solution through an expansion parameter $\epsilon$.  
\begin{align}
\eta&=\eta_{0}+\epsilon\eta_{1}+\epsilon^{2}\eta_{2}+\epsilon^{3}\eta_{3}\ldots\\
\Phi&=\Phi_{0}+\epsilon\Phi_{1}+\epsilon^{2}\Phi_{2}+\epsilon^{3}\Phi_{3}\ldots\\
\end{align}
Often $\epsilon$ is chosen as the ``wave slope'' $ka$ but sometimes other parameters are used.  The net momentum of the initial wave can be found by lagrangian methods or by simply computing the net mass flux above the line $z=-a$ since the motion below it is symmetrical.  The time averaged lagrangian flow gives an approximate mean  motion of
\begin{equation}
v_{d}=\omega k a^{2} e^{2kz}.
\end{equation}
The drift arises from the fact that the basis for the incompressible flow, $\nabla^{2}\Phi=0$, is given by $\sin(nx)e^{nz}$ and forwards motion requires the regions of forward motion and the crests of the waves be in phase.  Thus these regions always contain a net forwards momentum that is of order of the wave height and the net velocity of the particles in the motion.  This is seems to not be a result of a nonlinearity but incompressibility and irrotational motion since it arises at linear order.  As such, it seems unavoidable in any such solution and makes its absence in observations all the more puzzling.  

\begin{figure}
  \begin{centering}
 \includegraphics[width=2in,trim=20mm 220mm 10mm 20mm,clip]{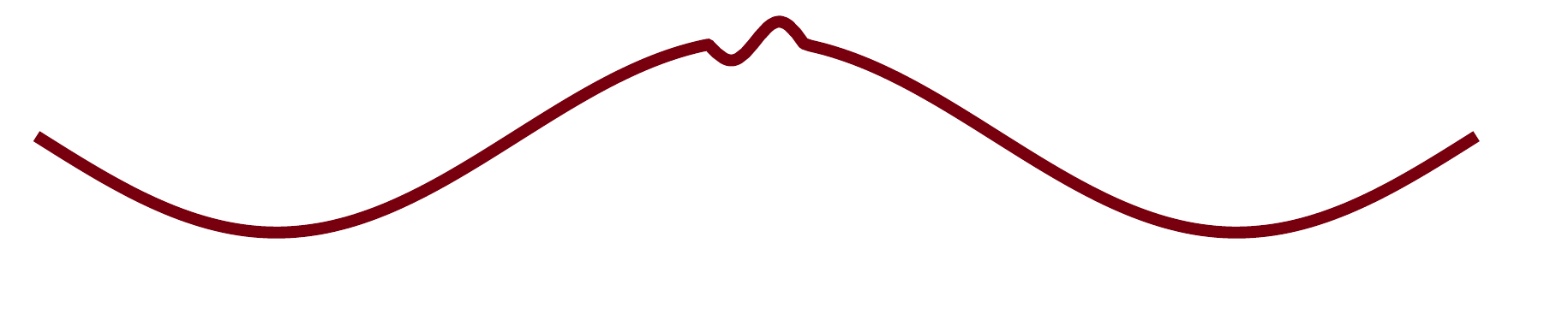}%
\hspace{1cm}  \includegraphics[width=2in,trim=20mm 220mm 10mm 20mm,clip]{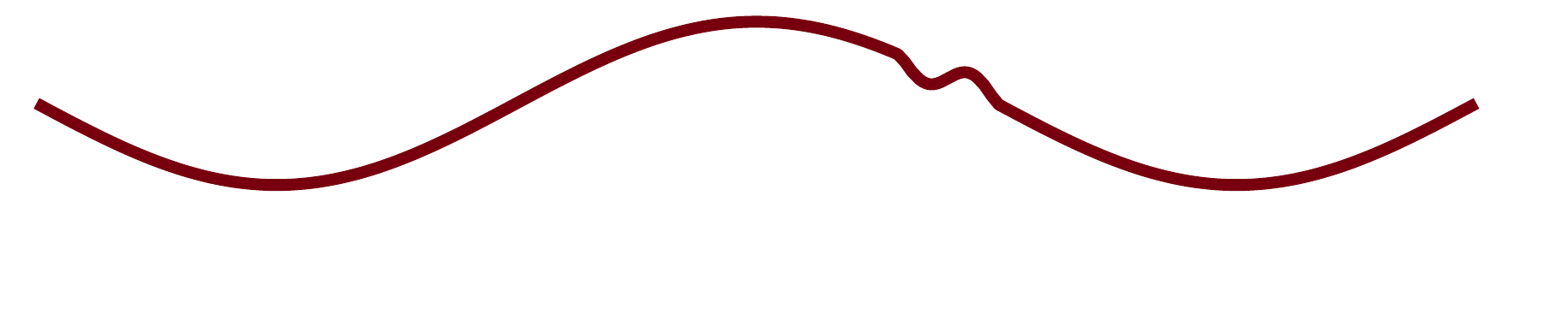}%
 \caption{\label{wavedouble} Two superpositions of ``small'' waves that do not give valid superposition of evolution over the period of the larger wave.  The mere shifting of the location of the small wave on the larger wave gives a large variation in the component wave decomposition of the system to conserve energy. }
 \end{centering}
 \end{figure}

Electromagnetic waves in media have a long history for illustrating the role of waves with damping, group velocity, causality and internally induced forces \cite{Chafin-em, Brillouin, Stokes}.  Their linearity has made them canonical examples for the study of wave motion.  Feynman called surface waves one of the worst possible examples \cite{Feynman} since ``everything can can go wrong does go wrong.''  Here he was referencing the effects of nonlinearity and surface instabilities.  To get a sense of how catastrophic small nonlinearities can be to superposition one need only consider the case of waves with wavelengths smaller than the wave amplitude of larger waves.  In particular, consider the relative amplitude and wavelength of each pair of waves as $(A,\Lambda)$ and $(a,\lambda)$ where $A\gg a$.  To be in the small wave limit described by Airy waves we must have $A\ll\Lambda$ and $a\ll\lambda$.  This is called the linear limit because the nonlinear term can be made arbitrarily small here.  However, to get a superposition where the N-S equations can decompose into independent equations of motion for each wave, we must not have $A\gg\lambda$.  Such a wave is illustrated in fig.\ \ref{wavedouble} in two different positions of relative phase.  This means that superposition is a reasonable supposition only in the case of a distribution of waves where the component waves do not vary over too many scales \cite{Chafin-windwaves}.  

One important distinction between acoustics and surface disturbances in fluids and electromagnetic waves in media is that the former are disturbances of the medium itself.  Electromagnetic waves can be microscopically decomposed from the medium so that the energy, momentum and motion can be uniquely partitioned between the two of them \cite{Chafin-em}.  Hidden phase shifts, temporal absorption and emission lags and backwards waves generate the internal forces in an intuitively clear and linear fashion.  In the case of medium disturbances, we must contend with the fact that flows and waves can exist simultaneously and interact.  This is technically possible for liquid dielectrics and EM waves but the relative time scales for these effects make them hard to realize in any detectable manner.  The Gerstner trochoidal wave can be thought of as a kind of perfect superposition of an irrotational wave and a shear flow.\footnote{It might be a reasonable definition of a ``shear flow'' to mean the purely vorticity dependent component of the flow as defined by the Helmholtz decomposition.} This allows every particle to move in a nonadvancing circular motion.  Certainly other shear flows are possible which let particles at different levels advance and retreat at varying net rates.   This is typically what is actually seen in observations.  

\subsection{Acoustic Waves}\label{Acoustic}

Acoustic waves are often assumed to have a net momentum at nonlinear order \cite{Lighthill,LL}.  The reason for this is that the eulerian infinite plane wave solutions give mass flow by lagrangian analysis.  The major weakness of this approach is that, as a result of looking at only infinite periodic solutions, we can't say anything about what the background flow is on top of them.  In the surface wave case, this was not a problem because the natural frame of reference was defined by the deep water velocity.  

The usual approach to this problem \cite{Lighthill, LL} is to find eulerian standing and traveling waves and then compute the net lagrangian drift of the particles.  The first approximation gives a periodic motion of the particle about its mean position and then higher corrections are derived.  This is frustrating since the drift then means the higher corrections leave us with no well defined mean position for higher corrections.  The drift seems to have nothing to do with nonlinear modification of the wave but the advective nature of the particle motion itself.  Instead of this, let us assume a solution where the particle motion is truly periodic and let the eulerian pressure and velocity information be modified.  

The particles can now be labeled with their respective mean positions $\bar x$.  Thus the particle position $p(\bar x,t)$ is a periodic function.  A first approximation is $p=\bar x+a\cos(k \bar x-\omega t)$.  Assuming the net displacement $a\ll1$ we have the approximate inverse $\bar x=p-a\cos(k \bar x-\omega t)$ where $\bar x(p)$ must be determined self consistently (iteratively).  The velocity of particle at $p$ is given by 
\begin{align}
u(p)=\frac{dp}{dt}&\approx\frac{d}{dt} (\bar x+a\cos(k \bar x-\omega t))\\
&\approx-a\omega\sin(k (\bar x+a\cos(k \bar x-\omega t) )-\omega t)\\
&\approx-a\omega\sin(k p-\omega t)-a^{2}k\omega\cos^{2}(k p-\omega t)
\end{align}
At any given instance, the positions $p$ are the locations where we need the velocities to give the eulerian description of the wave.  Taking a period cycled average we find $\langle u \rangle\approx-\frac{1}{2}a^{2}k\omega=-v_{d}$.  Thus the traveling wave solutions that move rightwards and have particles at rest have a net leftwards correction to their phase velocity of $v_{ph}=\frac{\omega}{k}-v_{d}$.  The implications of this will be discussed when we investigate packet motion.

\section{Periodic Solutions, Packets and Beams}

\subsection{Electromagnetic Waves}
The manner in which conservation laws are manifested in a dielectric medium has been in dispute since 1905 when the Abraham-Minkowskii debate began.  The momentum of an EM wave in a medium was deemed $p=n E/c$ by Minkowskii and $p=E/nc$ by Abraham \cite{Jackson}.  There have been many papers written on this subject \cite{Pfeifer} where some have even argued that the decomposition of electromagnetic and material momentum is not unique and a matter relegated to correct boundary conditions \cite{Pfeifer}.  Ironically, it is rather easy to create exactly solvable examples where this decomposition is unique and easily extracted \cite{Chafin-em}.  The source of the confusion (and doubt on this point) arises because the details of how forces arise at boundaries and at field gradients is typically not clearly distinguished.  By investigating microscopic models of media (where $\lambda\ll d$, the optical equilibration depth of the material) we arrive at a universal model of a dielectric packet as composed of both forwards and backwards moving waves.  The stress at the ends of the packet are given by this reflection in this packet gradient.  There is a second confounding effect that must be considered.  As the packet advances into the medium, the momentum seems to be microscopically manifested entirely by the EM fields.  The medium must ``ring up'' to a level of kinetic oscillation to become in equilibrium with the fields.  This absorbs both energy and momentum from the field and provides the outwards force on the medium spanning the packet explicitly.  The internal reflected field is not observed as an interfering field or standing wave due to a very slight continual phase shifting of it so that it is only seen as part of the advancing field.  This is carefully illustrated in a 2014 work by myself \cite{Chafin-em}.  At surfaces, the EM part of the entering beam is attenuated by the conversion to mechanical energy so gives an outwards force on the medium when all reflection is cancelled by antireflective layers.  By including the ``hidden'' backwards wave at these boundaries and packet gradients can we explain the direction of these forces microscopically.  

This example illustrates the importance of ``ring up'' effects\footnote{One can think of ring up as a hidden source of energy and momentum that is not evident from plane wave analyses.  In the hydrodynamic case, we will see that this energy and momentum can come from flows that are not even in the support of the waveforms.}  of waves at packet boundaries and gradients and, moreover, the importance of packet based analyses of waves compared to the unbounded periodic solutions we tend to favor for ease of integration.  The interaction of noninertial media with traversing waves such as accelerating dielectrics of flowing dielectric liquids is justifiably often neglected because the velocity changes over the equilibration times of the medium and EM waves are so small in practical cases.  However, they can potentially give cumulative effects and are interesting in that they give a case of wave-flow interaction where the waves and media can be decomposed into two different entities.  Because the waves are free waves between absorption and scattering events, this lends itself to an intuitive analysis.  As we will see below, when the waves and flow cannot be so microscopically decomposed, the situation is much more subtle and nonuniversal results arise that likely spell doom for the applicability of formal approaches like the generalized lagrangian mean (GLM).  

\subsection{Surface Waves}

There are three illustrative examples that help unravel paradoxical observations of waves and conserved quantities.  These are global periodic solutions, packets and beams.  The former are the usual cases but we have to make sure that perturbative corrections don't introduce hidden sources of vorticity or unphysical effects at infinity.  Two examples of such boundary effects at infinity are 1.\ the electromagnetic momentum of crossed E and B fields and 2. fictitious wavemaker forces.  In EM the problems arise so often that one should almost automatically assume that when you have created a system of such symmetry to make the integrals easy, that a finite fraction of the conserved quantities are hiding at infinity \cite{Chafin-em}.  The canceling electromagnetic momentum in stationary cases typically lives in fringe fields that vanish from our integrals in the case of infinite parallel plate fields.  In standard wavemaker theory, the momentum flux calculation neglects an essential integration constant that cannot be determined from the finite volume integrals computed \cite{Chafin-wavemaker}.  

In the case of beams of surface waves in a constant density fluid there will be some expected spreading that can be limited for beams broad compared to the wavelength.  The irrotational advance of motion in an incompressible fluid is only possible due to the time changing surface.   A beam of surface waves can then give an advance of fluid relative the deep wave and lateral fluid through superpositions and this mechanism without introducing vorticity.  A laterally localized acoustic beam in a fluid at rest does not have this property.  A point that is far from a surface with a time averaged advance relative to its neighbors means that vorticity must exist at the edge of the beam when the surrounding fluid is at rest.  We will return to this case when we consider acoustic waves.

Let us now consider packets of free surface waves.  It is problematic for a packet to advance at the wavespeed $u(k)$ which is much faster than the drift speed $v_{d}$ and give a net mass forwards.  To do so requires the packet have a net elevation since the fluid is incompressible.  Such a situation is not stable and the packet must descend to the equilibrium water level.  The net mass flux then would drive an elevation increase at the leading edge and a depression at the trailing one.  It has been suggested \cite{McIntyre} that this could lead to a deep irrotational recirculation underneath and around the edges of the packet to cancel the drift.  This gives a flow profile that is not observed.  If we could, by fiat of boundary conditions, enforce such a surface elevation change by elevating a region of sea or giving such a sloped profile then such a recirculation would be enforced by mass conservation and pressure gradients that extend to the boundary of the system.  However, given free surface boundary conditions we have the problem that the advancing mass inside the wave moves much slower than the wave itself.  The leading edge is driven by pressure that transmits motion at the speed of sound but pressure disturbances on their own impart no net forwards momentum to this parcel of fluid.  

I believe that the reason that there has been so much confusion on this topic, by myself and many others, is that surface waves are the least amenable to standard methods involving superposition methods of any waves.  The ability of surface instabilities to dominate the processes enforcing local conservation laws is the origin of this.  To understand why it is helpful to consider waves with more general surface conditions.  
Surface constrained waves (SCW), where we abandon free surface boundary conditions and the constant surface pressure constraint in favor of the kinds of changes we expect from superpositions of small waves, will generate completely different wave solutions than we physically observe.  These typically will have hidden tangential as well as normal and internal forces and so are not the kind of waves that arise from any reasonable pressure variation induced by the free surface.  Simultaneous momentum and mass conservation conditions require free surface waves have long range elevation changes beyond what the linear superposition can provide.  In contrast SCWs can generate long range deep irrotational flows and pressure fields that are not evident on the scale of wavelengths.  The length scales of packets and beams are often neglected in analyzing waves.  The following considerations will suggest that interacting propagating free waves typically cannot satisfy the verticality condition for $\eta(x)$.  Such instabilities provide a local avenue for wave to flow conversion.  

 \begin{figure}
  \begin{centering}
 \includegraphics[width=3in,trim=20mm 80mm 30mm 130mm,clip]{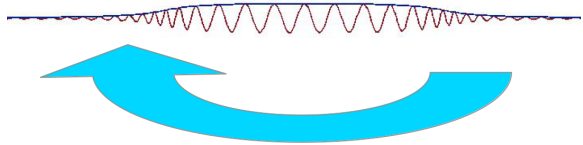}%
 \caption{\label{envelopewaveflow} For a rightwards propagating SCW solution, where only external surface forces are imposed, there must be an irrotational backflow to conserve mass.  The forces to generate this originate from packet scale pressure imbalances on the dynamic surface. }
 \end{centering}
 \end{figure} 

For a developing EM wave in a medium, there is a ring up time whereby some of the energy and pressure of the EM wave is absorbed by the packet at one end and returned at the trailing edge \cite{Chafin-em}.  The analogy for surface waves requires that the leading surface edge only transfer pressure and no net momentum - thus a standing wave contribution at the leading edge.  To balance momentum there must be a small reflected wave.  The stability of small waves traveling through a much larger wavetrain has been investigated \cite{Chafin-windwave} and found that they are unstable and must fail in microbreaking events.  Each advancing crest therefore must lose the equivalent momentum of its Stokes drift as it moves into undisturbed water.  One way to explain the relation between the group and phase velocity of deep water waves is in realizing that there is almost no kinetic energy flux and that all the energy is transferred forwards by the energy of the surface elevation \cite{Chafin-em}.  This potential energy flux then explains why wave packets appear as a cluster of waves that grow originating from the back and then diminish and vanish at the leading edge.  Indeed, one of the problems with a superposition over a broad range of frequencies is that, locally, the linear approximation is destined to fail badly with some regularity even for very small component wave mixtures.  

Returning to the Gerstner wave case, we can build a free surface packet of such waves but they do not persist as such.  The relatively slow backwards drift separates from the faster wave motion and we are left with advancing irrotational waves and a local shear flow that generates some packet scale elevation changes that carry off more energy as the vorticity diffuses into deeper regions of the fluid.  The advancing packet now has the problem that it is also generating some net motion of fluid in the opposite direction and so would seem to be leading to elevation changes over its length just as the shear flow.  However, this is our point of contention.  Let us consider an advancing packet of surface waves that is completely irrotational so that no internal shear flow exists.  In this picture, as wavelets rise up on the back end of the of packet they must pull mass into the wavetrain and then deposit it as they diminish at the other end. Since the wavespeed is so much faster than this velocity we argue that breaking occurs at the leading edge that cancels the net motion and conserves momentum \textit{locally}.  This is an energy dissipating process so that the wave is losing energy and vorticity is entering the flow from the free surface so that soon the trailing edge of the packet is moving through the backwards flow generated by the wave and a reverse breaking gives a momentum compensating flow as the packet advances through it.  First we will estimate the induced forces for SCWs then compare with results of this nonviscous damping of the packet due to this process.

 \begin{figure}
  \begin{centering}
 \includegraphics[width=3in,trim=20mm 80mm 30mm 70mm,clip]{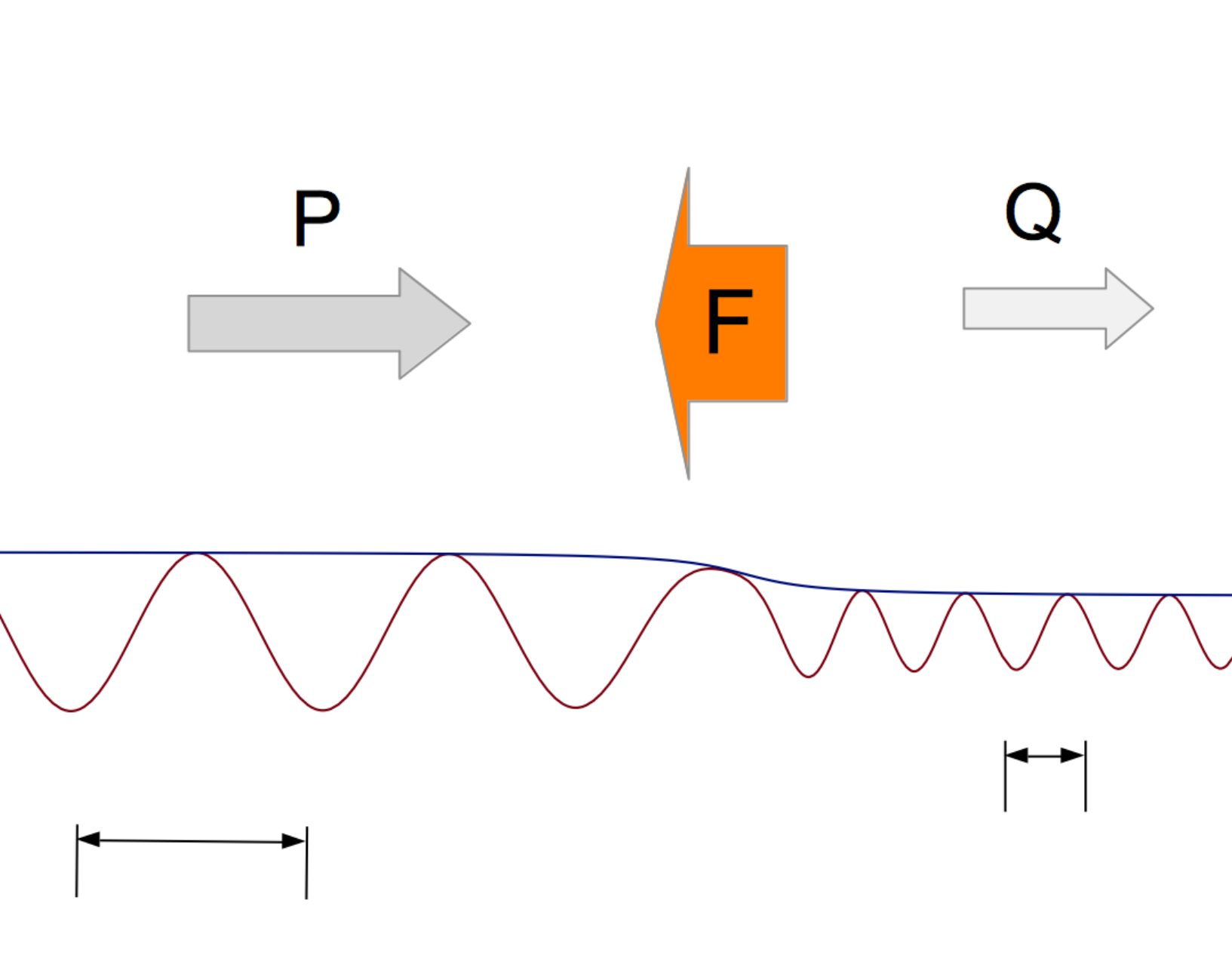}%
 \caption{\label{envelope} An advancing wave envelope joining two right moving surface wave forms. This is a (unphysical) surface constrained wave (SCW) solution where the fluxes are fixed to give local conservation of mass without larger scale flows. }
 \end{centering}
 \end{figure}

As a first example consider the SCW approach of an advancing envelope of a wave that is transitioning from amplitude and wavelength $(A,\Lambda)$ to $(a,\lambda)$ as in fig.\ \ref{envelope}.  As an infinite wave, we expect that sources and sinks over arbitrarily long distances are responsible for mass conservation at a depth not observable in the problem.  As in the case of free waves, we enforce that the envelope move at velocity of the potential energy flux, $v_{U}$.  The local waveforms on either side of the transition region advance at the free wave velocity $v_{ph}=\omega/k$ where $\omega^{2}=gk$ is the free wave dispersion relation.  When $\Lambda=\lambda$ we then have $v_{U}=v_{ph}=\frac{\omega}{k}=\frac{\Omega}{K}$.  Otherwise, the surface constraint themselves must provide potential energy flux to the surface.  In the free wave case, the wavelength is the natural length scale for waves to transmit information on the timescale of the period and, therefore, the scale on which we must seek local conservation of mass and momentum.  If the momentum flux density on the left of the transition is $P$ and the right is $Q$ then the force that must exist at the moving transition is $F=(P-Q)/d$ where $d$ is the transition length between the two waves.  This horizontal force is not just at the surface but must exist at depth to accelerate that entirety of the fluid involved in the wave motion.    This is a rather superficial example to show that forces are necessary at wave gradients and are not adequately imparted by the surface forces alone.  Now let us take a more thorough look at the conserved quantities of waves and their implications.  

The three conserved quantities we need to account for are mass, momentum and energy.  Angular momentum of waves exists but has subtle dependence on lateral boundary conditions so is not as interesting for a local analysis of conserved quantities.  The energy density (per area) is $E=\frac{1}{2}\rho g a^{2}$.  The momentum flux density is $f_{p}=\frac{1}{2}E$ so the momentum density (or mass flux density) is $p=f/v_{g}=E\frac{k}{\omega}=\frac{1}{2}\rho a^{2}\omega$.  Using that $v_{g}=\frac{1}{2}g/\omega$ and combining the fluxes together as a function of $a$ and $\omega$ we have
\begin{align}
f_{E}&=\frac{1}{4}\rho g^{2} \frac{a^{2}}{\omega}\\
f_{p}&=\frac{1}{4}\rho g a^{2}\\
f_{m}&=\frac{1}{2}\rho a^{2}\omega
\end{align}
To get a transition in amplitude of an incident wave described by $(A,\Omega)$ we have these three quantities to conserve.  To do so requires there be at least one reflected wave.  If we enforce SCW conditions there are surface forces to give deep mass flows.  For free surface waves, this is not an option.  Reflected modes are not observed in advancing wavepackets so we need another mechanism to conserve mass and momentum at wave gradients.

\subsubsection{Microbreaking}\label{breaking}

During my youth, I spent a great deal of time on the water rowing and kayaking along the coast of Maine.  One observation that always troubled me was how long glassy swells moving in to a channel would generate small sharp breaking actions along their faces, seemingly out of nowhere.  Now knowing that these waves had to shorten as they entered shallower water this suggests to me that these are essential momentum conserving features of waves that are dissipative and that, even for such ``small'' waves (i.e.\ ones with small slope or $ak$), the single valued assumption for the free surface $\eta(x)$ often fails.  Let us now consider the motion of a packet assuming that there is a microbreaking mechanism to cancel Stokes drift as it advances.  Without the surface constraints on motion above we have little choice in such a mechanism.  For incompressible fluids, vorticity can only enter through the surface by the vorticity transport theorem \cite{Batchelor,Lamb}.  

Consider a packet as in fig.\ \ref{envelopewaveflow} but without constraining surface forces.  The wavelength of the crests are assumed to be uniform so $\omega$ is constant.  As the advancing crests grow on the leading edge microbreaking events generate a backwards flow to generate mass conservation.  
The depth of the flow cannot initially be deeper than the length scale set by the difference in the wave heights of adjacent crests $\delta=a_{1}-a_{2}$.  The resulting shear flow profile then must begin at a depth on the order of $\delta$ deep.  The momentum density it must have (per area) is $\Delta p\approx-\rho  a\omega\delta$ so that the change in the velocity of the surface layer will be $\Delta v\approx-a\omega$.  This thin layer will slowly diffuse to a greater depth and lose its kinetic energy to heat.  

The energy lost (per volume) in this event from the wave is $-\Delta E\gtrsim\frac{1}{2}\rho (a\omega)^{2}=\frac{1}{2}\rho a^{2}\omega^{2}=\frac{1}{2}\rho g a^{2} k= E k$.  Conservation laws are built on eulerian not lagrangian quantities so, to get the power lost per area, we note that one wavelength ($\lambda=\frac{2\pi}{k}$) passes a perpendicular surface in time $\tau=\frac{2\pi}{\omega}$.  The rate of energy loss per area is then $|\dot{E}|\gtrsim  E k \delta \tau^{-1}= E k \hat{D}_{t}{\langle  a_{c}\rangle}=Ek \hat{D}_{x}\langle {a_{c}}\rangle v_{g}=\frac{1}{2} E \omega \hat{D}_{x}\langle {a_{c}}\rangle $ where $\langle {a_{c}}\rangle$ is the envelope or averaged crest height of the waves.  The trailing edge of the waves has a similar flux that leave no net horizontal momentum after the packet passes (although there will be some lateral shear at various depths depending on the rate that the fluxes have diffused).  One interesting result here is that there is no net momentum flux at all.  Since the packet builds up with canceling eulerian flows the local vertically averaged momentum density is zero.  Therefore there is no need to consider momentum flux balancing further in such an analysis.

\begin{figure}
  \begin{centering}
 \includegraphics[width=3in,trim=20mm 220mm 20mm 20mm,clip]{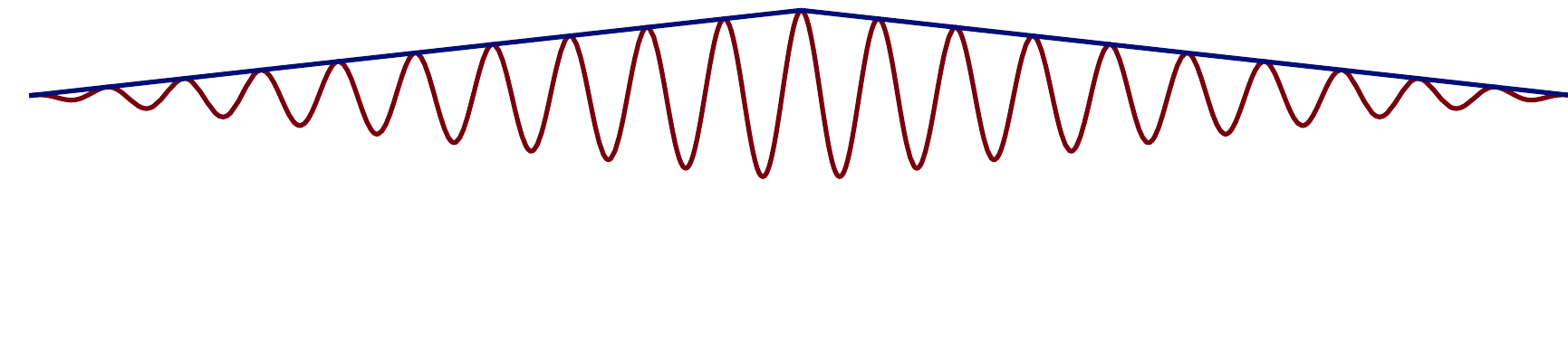}%
 \caption{\label{triangle} A triangular shaped wave packet of width $2L$ and maximum height $A$. }
 \end{centering}
 \end{figure}

These results now let us place bounds on the nonviscous attenuation of a packet.  For simplicity, let us consider a triangular shaped envelope of length $2L$ which has a maximum crest height of $A$ as in fig.\ \ref{triangle}.  Assuming the loss rate and duration of the experiment is small enough we don't need to consider shape changes in the packet.  We can compute the lower bound on the energy density loss rate to be
\begin{align}
\dot{E}=-E k \left(\frac{A}{L}\lambda\right) \tau^{-1}=-E \left(\frac{A}{L}\right)\omega
\end{align}
so that the amplitude of the $i$th crest varies as
\begin{align}
\frac{\dot{a}_{i}}{a_{i}}=-\frac{1}{2}\left(\frac{A}{L}\right)\omega
\end{align}
Defining the slope of the packet from the crests to be $m=A/L$ and letting $l_{i}$ be the distance from the edge of the packet to the $i$th crest with amplitude $a_{i}$ we have
\begin{align}
\dot{m}=-\left(\frac{\omega}{2}\right) m^{2}.
\end{align}
The slope of the packet then evolves as
\begin{align}
m(t)=\frac{1}{\frac{L}{A}+\frac{\omega}{2}t}
\end{align}
so that the packet has an amplitude half-life of $\tau_{1/2}=\frac{2L}{A\omega}$.  The viscous loss of energy per volume is $\frac{1}{2}\mu \nabla v\colon\nabla v$ so the power lost per area is
$\frac{1}{4}\mu a^{2}\omega^{2} k=\frac{1}{4}\mu a^{2} g k^{2}=\frac{1}{2}E \frac{\mu}{\rho} k^{2}=\frac{1}{2}E \nu k^{2}$.  Comparing this with $\frac{1}{2} E \omega \hat{D}_{x}\langle {a_{c}}\rangle$ we see that the ratio of viscous to nonviscous damping is
\begin{align}
\zeta=\frac{\nu k^{2}}{\omega \hat{D}_{x}\langle {a_{c}}\rangle}=\frac{\nu k/v_{ph}}{m}.
\end{align}
For water $\nu\sim10^{-3}$m$^{2}$/s.  For ocean waves, typical values for $k$ and $v_{ph}$ are such that $k/v_{ph}=\omega^{3}/g^{2}\gtrsim10^{-2}$ so for packets to damp faster from viscous losses than microbreaking the slopes must be such that $m\lesssim10^{-5}$.  

We can investigate the case of a square packet an the nonviscous damping.  In this case, the end waves make all the loss contributions and give net initial energy change $2 E k a \tau^{-1}$.  This is the same total loss rate as above.  Unlike the triangular wave packet, here all the losses are at the ends and the packet shape gets eroded.  However this illustrates a general features of such monochromatic packets.  The total loss is a function of the oscillation of the envelope, specifically the value
\begin{align}
\textit{o}=\int | \hat{D}_{x}\langle {a_{c}}\rangle | dx
\end{align}    
determines the loss rate by $\frac{1}{2} E \omega \textit{o}$ as the energy loss of the total packet.  Thus the momentum flux conserving losses are not a generally linear effect but do give a universal loss rate mechanism.  

An idealize wavemaker analysis that imparts and removes mass flows at the ends does give true Stokes drift \cite{Chafin-wavemaker}.  From this the strong bounds on impulses of wavemakers can be established.  However, given that waves are generated in such systems by flappers and that waves must cross the tanks that don't have Gerstner-like imparted flows suggests microbreaking will again be at work.  The fact that these tanks tend to be closed at both ends implies that the net flow will be zero regardless of the size of this effect.  Even tanks that allow flow at the wavebreaks at the end will generally stop the diffusing vorticity layer carrying mass below this height so are not likely to observe true Stokes drift.  Given the relatively short length of such tanks compared to open bodies of water, the vorticity generated by wave breaking at the tank ends has the potential to generate deep currents that are dependent on tank shape and give surface flows unrelated to Stokes drift \cite{Bagnold}.

\subsection{Acoustic Waves}
The most dramatic example of the interaction of waves and flows is that of acoustic or Eckart streaming.  In this case MHz frequency sound beams can drive fast streams of fluid away from the driver.  It is known that Stokes drift is far too small to account for this effect but it is often brought up as a relevant source of momentum to be accounted for along with Reynolds stresses.  I will show both of these are largely unimportant.  The discussion will investigate several fundamental questions.  Firstly, is there an analog to the SCW approach to acoustic waves?  In what ways is surface wave Stokes drift related to that of acoustic waves.  In the surface wave case we found conditions on monochromatic waves that are in the linear limit but violate this in superposition.  Are there similar superposition failure regimes for acoustic waves?  Finally, I will investigate the kinds of forces and contributions Reynolds stress can and cannot provide in fluids.  Most importantly, the role of energy stored in the medium outside the support of packets and beams will be shown to often be comparable or larger than the energy in the sound itself.  Such a state suggests that superposition is likely to fail even in cases where the naive linear limit seems valid and that infinite periodic solutions are lacking essential information that has been pushed off to infinity.  

\subsubsection{Sonic Wavemaker}

Let us first consider the implications of packet and beam decompositions of waves on the Stokes drift we traditionally associate with sound waves.  Consider a long broad tank of fluid with a beam of sound that can be driven through it and controlled at will.  The driver can further be isolated from the fluid by a membrane through which the sound passes so no vorticity generated at the driver can migrate through this part of the fluid.  Surface waves allow incompressible fluid to transport mass irrotationally because of the dynamic free surface.  For acoustic waves, even though the medium is not truly incompressible, there is no such mechanism since the mean local density never changes.  Let us consider an idealized sonic wavemaker to investigate mass flux and sound.  This is in direct analogy with the surface wavemaker that I presented recently to investigate the kinds of stresses that can be induced by such devices \cite{Chafin-wavemaker}.  

 \begin{figure}
  \begin{centering}
 \includegraphics[width=3in,trim=10mm 120mm 10mm 65mm,clip]{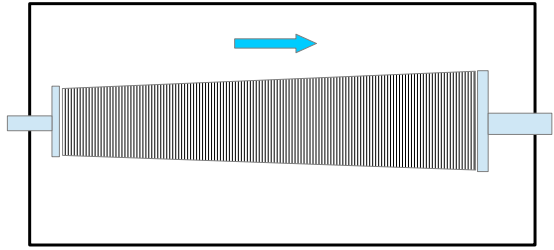}%
 \caption{\label{beam} Oscillating wave driver and absorber in a closed volume of fluid.}
 \end{centering}
 \end{figure}

For planar acoustic waves, it is ambiguous if there is an underlying parallel flow of the fluid.  Unlike surface waves, there is no way to decompose the flow from acoustic motion other than integrating out the net lagrangian motion.  By using packets and beams we can, however, gain some further insights.  Consider the case of an acoustic beam driven in a chamber as in fig.\ \ref{beam} where we carefully drive the actuators to create a rightwards traveling wave between them.  If there were some net momentum to the beam we would have a problem at the ends of the drivers where a mass flux would have to be conserved.  This is the reason that when we investigated the drift in sec.\ \ref{Acoustic} that we insisted the basis solutions be cases where the net particle motion was periodic so the mean mass flux was zero.  

  \begin{figure}
  \begin{centering}
 \includegraphics[width=3in,trim=10mm 120mm 10mm 65mm,clip]{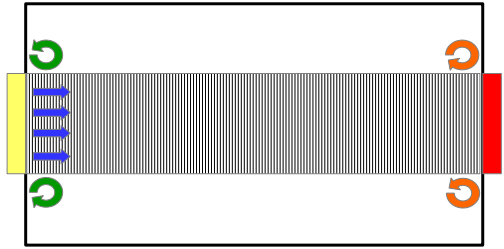}%
 \caption{\label{sonicwavemaker} A generalized wavemaker that can generate arbitrary acoustic waves and parallel flows simultaneously.}
 \end{centering}
 \end{figure}
 
Let us consider a sort of maximally generalized version of an acoustic generator analogous to the surface wavemakers in \cite{Chafin-wavemaker} as in fig.\ \ref{sonicwavemaker}.  This is a wavemaker in an inviscid fluid with drivers and absorbers that can be tailored to impart and absorb any wavemotion and to impart any net lagrangian flow along with the waves.  We have assumed the fluid is inviscid and suffer no thermally diffusive losses and the wavelengths generated are narrow enough that the beam spreading is negligible.  Note that the width of the chamber can be chosen arbitrarily thin, even less than one wavelength as long as there is no laterally confining geometry introduced by the chamber.  

When we set up a rightwards propagating wave with a net lagrangian motion of velocity $v$ (not necessarily equal to the Stokes drift) we introduce a singular source of vorticity at the corners of the driver that propagate with velocity $v/2$ and a corresponding sink at the absorber.  Notice that this drift can actually be established in opposite direction to the wave motion.  Excluding the lateral effects from this vorticity line and the support of the beam we have a basis of solutions that can be characterized by $k$, the amplitude $a$ and the lagrangian drift $v$.  The frequency $\omega$ is a function of these.  If we use the eulerian point of view and neglect the drift the dispersion relation is $v_{ph}=\omega/k$ as a constant until we reach high enough frequencies for nonlinearities to enter.  Amorphous solids and liquids obey this result very well over large ranges in frequency.  If we introduce some additional rightwards velocity shift of the medium $u$ then the dispersion relation becomes asymmetrical so that standing and right and left traveling waves now give different results.  

\begin{figure}
  \begin{centering}
 \includegraphics[width=3in,trim=10mm 120mm 0mm 65mm,clip]{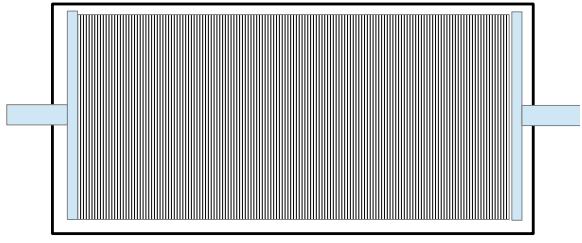}%
 \caption{\label{beamequal} An acoustic wavemaker with parallel wavefronts and no mass or momentum sourcing at the driver plates.}
 \end{centering}
 \end{figure}

The new frequency for the right moving waves is $\omega=(v_{ph}+u)k$.\footnote{The drift is a function of the frequency $\omega$ so there is some self consistency that will need to be considered shortly.}  This seems like an insubstantial and trivial result however remember that this $u$ is not the lagrangian drift.  The rate of net mass transport is $u+v_{d}$.  If we are to consider a wavemaker that is unlike that of fig.\ \ref{sonicwavemaker} and more simply and realistically like that driven by plates and absorbers as in fig.\ \ref{beamequal} then the physical basis set we care about are the states with no averaged net drift.  When we consider the case of beams with lateral confinement of the acoustic motion then there is an evident problem with these distinct flows.  They each give different vorticity at the edges of the beams.  Vorticity is an advected quantity and is only created in particular fashions by baroclinic means.  This means that simply changing an acoustic field, which can be done very rapidly, cannot create a meaningful change in vorticity.  In the following, vorticity will play a further important role in distinguishing physical mechanisms of acoustic forces on flows and general wave-flow interactions.  

\subsubsection{Dispersion and Superposition Limits}
The drift motion of the underlying flow will leave the wavelength fixed but alters the frequency that the waves exhibit as measured by an eulerian (fixed background) observer.  This will lead to the somewhat unnerving result that the dispersion relations will be different for standing and translational waves in blatant contradiction with the idea of linear theory for which dispersion relations are relevant.  We will see that, suitably restricted, such relations still have some value but, just as we saw in the surface wave case, there are superpositions of monochromatic waves that are linearly valid but that do not evolve as linear superpositions over short timescales.  

For \textit{standing waves} there is no drift correction so the dispersion relation is $\omega=(c+|v|)k$ where $c$ is the intrinsic speed of sound of the material and $v$ is the translational velocity of the medium.  If we combine such solutions with the same $v$ then there are no drift corrections and superposition is valid in that we produce a new ``small'' wave solution and it evolves as the sum of the monochromatic components for long times i.e.\ given an initial standing wave $f(x)=\sum_{i}^{n}a_{i}\cos(k x)$ 
\begin{align}\label{super}
f(x,t)=\sum_{i}^{n}a_{i}\cos(k x- v t)\sin(\omega t)
\end{align}
for time scale of many oscillation periods when the $a_{i}\ll k^{-1}$ and as long as the net amplitude maxima are always smaller than the local gradients i.e.\  
\begin{align}
|\nabla f|\ll1.  
\end{align}
This first statement is a standard result but the second is important to consider especially if we have a broad distribution of frequencies so that arbitrarily large amplitude oscillations can eventually be created at some location.  If we add two such waves at different $v$ then we have the strange situation of adding two standing waves and two different flow velocities.  Let us now consider the case of adding two standing wave solutions on oppositely moving flows $(a,k,v)$ and $(a,k',-v)$, where we have specified the solutions by giving the amplitude, wavevector and net flow drift.  That the solutions are standing waves when boosted to the rest velocity of the flows is assumed.  
The net velocity of the resulting fluid must be zero and the wave solutions would have to be propagating waves if the amplitudes were linear superpositions over time.  However, the pressure fields that drive a standing and propagating wave are out of phase by $\pi/2$ so adding the velocity and pressure field information cannot give the required pressure field for a propagating wave.  Thus superposition for waves with different underlying flows gives an evolution that does not satisfy the analog of eqn.\ \ref{super}.  Because of this we only consider wave superpositions that have the same underlying drift.  On its face, this is not a severe restriction since we tend to think of waves as disturbances on top of an underlying medium and superimposing media of different states is an idea that, outside of quantum mechanics, seems rather artificial.  

Let us now consider the case of propagating waves.  The boundary conditions for a medium in a fixed box at rest and our analysis of laterally localized beams suggests that vorticity constraints require that no net flow be present.  We must allow some small oscillatory behavior at the boundaries to drive a beam and absorb it but no matter flux can pass through this surface.  Let us now construct a basis of propagating monochromatic waves that obey the condition of zero local net flux.  Based on our above notation we are then interested in the right moving waves $(a,k,-v_{d})$.  Since $v_{d}$ is a function of the frequency we must find such a consistent set.  Since the drift speeds are generally much less than the wave speed, to lowest order we can ignore any translational correction to the frequency and use the basis of functions found in sec.\ \ref{Acoustic} to give the eulerian velocity field of a drift free wave:
\begin{align}
u(x,t)=a\omega\cos(k x-(\omega-v_{d}k) t)
\end{align}
where $v_{d}=\frac{1}{2}a^{2}k\omega$.  Note that this depends in a nonlinear way on the amplitude itself so even superimposing two identical waves is problematic.  The resulting ``dispersion relations'' uses the standing wave $\omega=ck$ as a lowest approximation to give $v_{ph}=c (1 - \frac{1}{2}a^{2}k^{2})$
so that the corrected relation is
\begin{align}
\omega=c k (1 - \frac{1}{2}a^{2}k^{2})
\end{align}
The left moving waves will have a similar correction but there is no way to combine such right and left moving waves of the same amplitude to get a standing wave.  The relevant standing waves for our cavity have dispersion relation $\omega = c k$ so that the frequency for the same wavelength is larger.  This suggests that two oppositely moving packets will give essential scattering and frequency mixing.  Interestingly, for a fixed amplitude, this gives a wave where $v_{g}=\partial \omega/\partial k< v_{ph}=\omega/k$.  Longer wavelengths have a decrease in oscillation because the of a slower phase velocity for the same kinetic energy.  This results in a lower energy flux for a given energy density.

These functions are the ``best possible'' basis for our cavity since they are the only reasonable functions to be combined into packets subject to the mass flux conservation condition at the boundaries.  Sadly, these dispersion relations are essentially amplitude dependent as well as a function of the net direction of motion.  In this sense, acoustic waves, just like surface waves, by virtue of being derived from hydrodynamics, are \textit{essentially} nonlinear.  There is no linear limit in general for superpositions of small waves.  The usual condition on the relative bounds on wavelengths and amplitudes for a pairwise superposition of waves still is of concern here but it is probably more interesting to consider just what sort of packet dynamics exist.  

\subsubsection{Packets}
There is no analog of surface wave breaking for packets therefore energy must be conserved at the level of zero viscous dissipation.  We have already argued that a sound wave moving into a resting medium will not generate any net motion except for possibly at the laterally attenuated regions.  Even in this case, wave energy attenuation is probably necessary.  This means that the time averaged momentum density on the time scale of oscillations will vanish and energy flux is our main concern.  Let us begin with some idealized cases where fictitious external forces enforce ideal behavior analogous to the surface constraining forces for surface waves.  

\begin{figure}
  \begin{centering}
 \includegraphics[width=3in,trim=10mm 120mm 0mm 65mm,clip]{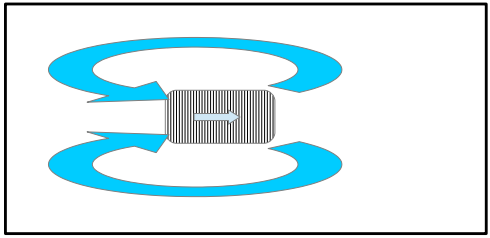}%
 \caption{\label{packetflow} A monochromatic eulerian solution packet constrained in its shape by external forces in a closed tank.  The induced backflow conserves mass but also hides a source of energy and momentum not in the support of the packet. }
 \end{centering}
 \end{figure}

As a first trial let us consider a tapered monochromatic packet built of superimposed eulerian traveling solutions that advances rightwards at $c$ and imposes external forces required to keep it in this shape as in fig.\ \ref{packetflow}.  This solution gives a mass flux which must recirculate at the ends to conserve mass as in the case of SCW solutions to surface waves.  As the packet moves to different locations relative to the walls, these may absorb some forces that are then required to be imposed on the packet.  The packet presumably is created from the fluid which is initially at rest.  The backflow introduces an energy that is not in the support of the packet so gives a hidden ``ring up'' energy analogous to the energy sink of the medium as an electromagnetic packet advances into a dielectric \cite{Chafin-em}.  At the edges of the packet there is a time averaged singular vorticity analogous to that generated by the general wavemaker in fig.\ \ref{sonicwavemaker} indicated that these external forces generate sources and sinks of vorticity as well as confine the motion of the packet and necessary end-of-packet forces for it to evolve as such.    

\begin{figure}
  \begin{centering}
 \includegraphics[width=3in,trim=30mm 210mm 30mm 10mm,clip]{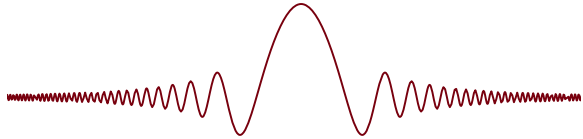}%
 \caption{\label{stationarypacket} A shape preserving translating packet. }
 \end{centering}
 \end{figure}

Let us now seek packet solutions that do not depend on external forces.  In this case we expect a zero locally average momentum density and an energy flux that is consistent with the packet motion.  Since we expect no motion outside the support of the packet, this means that there must be backwards drift sized corrections to the advancing component waves in the packet.  For a given envelope we seek a condition on the frequency variations needed to make such an envelope stationary in shape.  Consider a wave envelope $f=A(1+x^{2})^{-1}$
 built out of our zero mass flux rightwards acoustic waves.  Instead of taking superpositions, which we have seen is problematic, let us specify a waveform where the amplitude and wavelength varies slowly.  To get a stationary solution with constant $\omega$ we need the phase velocity to be constant at all parts of the wave.  The phase velocity is
\begin{align}
v_{ph}=c\left(1-\frac{1}{2}a^{2}k^{2}\right),
\end{align}
so that, if the amplitude crests follow the envelope function $f(x)$ at $t=0$, we can assign $k(x)=\langle a_{c}\rangle^{-1}$ to get $v_{ph}$ to be constant.  The resulting packet is illustrated in fig.\ \ref{stationarypacket}.  
There is a problem of unbounded oscillations at the tails of the packet.  Realistic packets would need a frequency cutoff so it it cannot be of infinite lifetime.  
In a sense it is ``monochromatic'' since the frequency of oscillation is constant, however, it certainly not constant wavelength.  Since physical disturbances causing sound are often more likely to produce such constant wavelength packets we should enquire as to their fate.  The peak of such a packet will travel slowest while the leading edge advances away.  The trailing edge will tend to catch up to the center until the wavelengths shorten so as to limit this behavior.  It is often stated that nonlinearities will lead to sound condensing into a limiting sequence of shocks.  This seems inconsistent with the behavior of such packets so the topic may warrant further consideration.

\subsubsection{Reynolds Stesses}
Reynolds stress arises in the time averaging of periodic and turbulent flows.  The time averaging of the flow in an incompressible fluid gives
\begin{align}
\rho\left(\partial_{t}\bar u+\bar u \cdot \nabla \bar u  \right)=-\nabla \bar p+\nabla\cdot\left(\mu \nabla \bar u-\rho \overline{u\otimes u}  \right)
\end{align}
The last term, $\rho \overline{u\otimes u}=\rho u_{i}u_{j}$, is the ``Reynolds stress.''  Its structure is interesting because it looks like a flux of  $i$-momentum across the $j$th face but contains no viscosity coefficient so is a purely kinetic quantity.  To generate parallel plate-type shear flow we typically need a baroclinic source but there is none for such a fluid.  One can have extension flows where there is no apparent vorticity creation (though often this is just pushed off to infinity or a boundary).  For acoustic streaming we need the former type of shear motion with vorticity generation.  

Reynolds averaged stresses are typically used in approaches to turbulence where evolving the stresses are a kind of closure problem with the ultimate goal of self consistency.  Acoustic streaming experiments are often largely free of turbulence with large steady vortical flows as the stationary result.  The above equations are not entirely applicable to acoustic streaming because acoustics requires some small compressibility.  However, we now have seen how beams in fluids with fixed simple boundary conditions and of a trivial topology don't have propagating solutions that result in a net mass motion.  This is not to say that eventually such a flow could build up from some attenuation at the beam edge but acoustic beams can be turned on and off extraordinarily fast.  These streaming flows take some time to establish.  This is a reason to doubt naive and formal uses of Reynolds stress in generating flow.  

\begin{figure}
  \begin{centering}
 \includegraphics[width=3in,trim=10mm 120mm 0mm 20mm,clip]{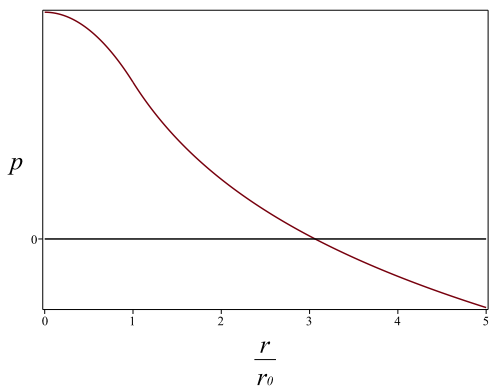}%
 \caption{\label{pressureplot} The Reynolds induced pressure for a cylindrical beam in a cylindrical container.  The container radius is chosen to be $R=5r_{0}$. }
 \end{centering}
 \end{figure}

As an example of the way these advective nonlinear terms can alter the forces on the fluid consider a stationary traveling beam in a cylindrical chamber as in fig.\ \ref{beam} however let the length of the chamber, $L$, be comparable to or less than the beam radius, $r_{0}$, and the wavelength much less than the beam radius so that spreading is small over the length of the chamber.  Let the chamber be of radius, $R$.  Since the beam is steady and the oscillations are very rapid compared to the acoustic crossing time for the beam, we can give an equation for the ``coarse grained'' pressure based on time averaging of the Reynolds stress:
\begin{align}
\nabla^{2}\langle p\rangle=-\rho\langle |\nabla v|^{2}  \rangle
\end{align}
to lowest order in $a k$.  Because we have a fluid at rest that has then had an oscillatory beam introduced we have assumed the motion is irrotational.  Here the averaging is over time scales of sound oscillation.  For our uniform beam of wavevector $k$ and amplitude $a$ this gives a static Poisson equation for the pressure field away from the beam $\nabla^{2}\langle p\rangle=-\frac{1}{2}\rho (a k\omega)^{2}\Theta(r_{0}-r)$.  The solution inside and outside the support of the beam is 
\begin{align}
\langle p\rangle=
  \begin{cases}
   p_{0} -  \frac{\rho (a k\omega)^{2}}{8} r^{2}      & r<r_{0}\\
   p_{1} \ln(r/r_{0})  +p_{2}      & \text{if } r \geq r_{0} 
  \end{cases}
\end{align}
Matching the pressures and gradients of pressure at $r=r_{0}$, the resulting equations are
\begin{align}\label{pressure}
\langle p\rangle=
  \begin{cases}
   p_{0} -  \frac{\rho (a k\omega)^{2}}{8} r^{2}      & r<r_{0}\\
   p_{0}-\frac{1}{4}\rho (a k\omega)^{2} r_{0}^{2}\left( \frac{1}{2}+\ln\left( \frac{r}{r_{0}} \right) \right)     & \text{if } r \geq r_{0} 
  \end{cases}
\end{align}
where we can derive $p_{0}$ from the mass conservation of fluid in the medium.  We should remember that, at constant $T$ we can define $p(\rho)$ so the above equations only make sense to the extent that variations in the density from changes in $p$ are relatively small.  
The isothermal compressibility, defined by 
\begin{align}
\beta=\frac{1}{\rho}\frac{\partial \rho}{\partial p}
\end{align}
is a very small number for liquids.  Let $\rho(p=0)=\rho_{0}$ so the small change in density due to pressure is $\Delta \rho=\rho(1+\beta p)$.  Mass conservation then requires that $2\pi\int p(r) r dr=0$ so the central pressure is
\begin{align}
p_{0}=\frac{1}{16}\rho_{0}(a k\omega)^{2}r_{0}^{2}\left( 4\ln\left(\frac{R}{r_{0}} \right)+\frac{r_{0}^{2}}{R^{2}}  \right)
\end{align}
As long as $\beta p_{0}\ll1$, eqn.\ \ref{pressure} is a valid expression.  The pressure profile is plotted in fig.\ \ref{pressureplot} with a beam radius to container radius ratio of $r_{0}/R=5$.  

The role of the Reynolds ``stresses'' in this situation seems to be simply to generate an extra density in the support of the acoustic beam that will then be immediately cancelled by an elastic shift of the matter outwards so that $\langle p\rangle$ is constant.  The resulting coarse grained equilibrium density is $\langle \rho(r)\rangle=\rho_{0}(1-\beta\langle p\rangle)$.  
This is the extent of mass redistribution produced by the acoustic field.  The logarithmic dependence on the pressure correction suggests that, while such effects are typically small, the formal limit of mass redistribution in an infinitely wide chamber is unbounded.  (For a 2D sheet acoustic beam the effects would not fall off even logarithmically so give even larger effects.)  

The energy of compression and mass displacement are $\sim \beta \rho^{2} (a k\omega)^{4}r_{0}^{4}(R^{2}L)$.  This sets a relative energy scale of the elastic response to the acoustic beam energy as
\begin{align}
\frac{E_{el}}{E_{ac}}\sim\beta\rho(a\omega)^{2}(k r_{0})^{4}
\end{align}
While $\beta\sim10^{-10}$~Pa$^{-1}$ and the particle velocity $a\omega$ is typically very small, the width of the beam in wavelengths, $r_{0}/\lambda$, is generally very large.  
The importance of this is that $E_{el}$ represents energy associated with the propagating beam that is not in the support of it.  If we were to have a localized packet, the energy would be confined to an integrable volume regardless of the chamber size but it is still largely external to the packet as mass must get pushed out of the support and then recontracted as the packet passes.  We see that this can become a rather large fraction of the system energy even when we are well within the linear regime of our approximations for $p(\rho)$ corrections.  If we were to consider a transitional state of a beam with a leading edge advancing across the cavity, this elastic and kinetic energy of the beam must be established at a speed comparable to $c$, the speed of sound in the medium.  It is telling that the Reynolds ``stress'' does not exhibit any effect but a pressure and elastic correction so generates no net motion of the fluid along the beam length once transients have dissipated.

\section{Conclusions}
The excitations of hydrodynamics have been shown to have fundamental nonlinearities that can invalidate superposition on surprisingly short time scales.  The boundary effects of infinite solutions often obscures this but by using packets and beams this becomes more evident.  Any question involving mass drift is at a level where nonlinear corrections must be included.  The dramatic consequences of this are that packets of surface waves generate continuous microbreaking and nonviscous wave damping and that acoustic waves have a retarded correction to the phase velocity for traveling waves that prevent any time averaged momentum density from arising.  An extreme example is the case of the superposition of waves of very uneven sizes which display rapid breakdown of linear superposition over very short time scales.  The relevant dispersion relation for acoustic waves, at the level where mass drift matters, must bifurcate into distinct cases dependent on whether they are standing or propagating.  Only very particularly tailored monochromatic subsets of these can be superimposed to generate waves of a single propagated wavevector.  The resulting effects for both cases is for Stokes drift to be cancelled.  

There are interesting auxiliary features that arise from these corrections and the Reynolds stress terms.  These typically don't drive any mass flux with the wave but can give lateral displacements of fluid outside the support of the beams that can easily be larger than the energy in the beam itself.  It is possible that such contributions are typical of situations where superposition of linearized waves fails for components that are well within the linear limit.  These ``hidden'' sources of energy give reason to reconsider local wave interaction models and generalized lagrangian mean (GLM) approaches to wave interactions.  The vanishing of net local momentum density simplifies conservation law considerations and greatly constrains evolution and damping of these systems.  In the case of surface waves, the interactions of the waves with the flow are so strong that it makes questionable the value of building Fourier decompositions of wave motion since these will tend to change rapidly, even on the scale of a single wave period \cite{Ha}.  

Future work should be an experimental investigation into microbreaking and wave damping and a theoretical investigation into acoustic streaming given the constraints on wave structure and forces expressed in this paper.  The appearance of nonviscous damping mechanisms for mass flux balancing of surface waves may suggest that analogous considerations are needed to understand acoustic streaming.  This article is meant as a preparatory work that gives a serious reconsideration of the theory of wave-flow interactions.  These are clearly essential in understanding heat and flow interactions since sound is merely a coherent version of the quantum excitations of thermal phonon excitations and have important consequences for the extreme shears that can occur in microfluidics.

\end{document}